\documentclass[onecolumn]{revtex4}

\usepackage{graphicx}% Include figure files
\usepackage{dcolumn}% Align table columns on decimal point
\usepackage{color}
\usepackage{bm}% bold math

\input epsf

\begin{document}

\title{\Large Radion cosmology and stabilization}

\author{\bf Sumanta
Chakraborty 
\footnote{sumantac.physics@gmail.com}}

\affiliation{IUCAA, Post Bag 4, Ganeshkhind,
Pune University Campus, 
Pune 411 007, India}

\author{\bf Soumitra SenGupta \footnote{tpssg@iacs.res.in}}

\affiliation{Department of Theoretical Physics,
Indian Association for the Cultivation of Science,
Kolkata-700032, India}

\date{\today}

%%%%%%%%%%%%%%%%%%%%%%%%%%%%%
\begin{abstract}
We solve the Einstein's equation in five-dimensional spacetime for Randall Sundrum Brane world 
model with time dependent radion field to study variation of brane scale factor 
with time. We have shown that as radion field decreases with time compactifying 
the extra dimension, the scale factor increases exponentially with time leading 
to inflationary scenario. We have also proposed a time dependent 
generalization of Goldberger-Wise moduli stabilization mechanism to  
explain the time evolution of the radion field to reach a stable value after 
which the scale factor on the brane exits from inflationary expansion.
\end{abstract}
%%%%%%%%%%%%%%%%%%%%%%%%%%%%%%%

\maketitle

%%%%%%%%%%%%%%%%%%%%%%%%%%%%%%%%%%%%%%%%%%%%%%%%%%%%%%%%
%%%%%%%%%%%%%%%%%%%%%%%%%%%%%%%%%%%%%%%%%%%%%
\section{Introduction}\label{braneintro}
The apparent mismatch between the fundamental scales of particle
physics and gravity is resolved in recent years by changing the
behavior of gravity at short distance. One of the most promising
candidate for resolving this is introduction of large extra
dimension and reducing the scale of quantum gravity all way down
to weak scale which was shown first by Arkani-Hamed \textit{et al.}
\cite{Hamed1,Antoniadis,Horava,Giddings}. This
possibility might imply the existence of large extra dimensions
and this opens up new cosmological scenario for the early universe
\cite{Lukas,Hamed2,Dienes}. Though an
interesting concept, the existence of a large hierarchy between
the weak and the Planck scale leads to new hierarchy through the
length scale of large extra dimension where the radius of extra
dimension is much large than natural value.

We could have a different setup where the extra dimensions are small 
but the background metric is not flat along extra co-ordinate.
This possibility was first analyzed by Randall and Sundrum (RS) \cite{Randall1,Randall2}.
The metric along the extra dimension was a slice of anti-de Sitter 
(AdS$_{5}$) space due to negative cosmological constant balanced by two
brane tensions. This non flat geometry causes physical scales on 
two branes to be different and exponentially suppressed on the negative tension
brane. Some generalizations of the RS models are discussed in works 
by Goldberger and Wise \cite{Wise1,Hamed3,Oda,Kraus,sdssg}
and embedding this into supergravity has been discussed by 
Hawking \textit{et al.} \cite{Hawking,DeWolfe}.

The cosmology of this model can be very different from the
ordinary inflationary cosmology in four dimension. The study of
early cosmology in RS brane world is appeared to be hindered by
one obstacle i.e. the late time cosmology differs widely from the
usual Friedman-Robertson-Walker (FRW) universe in the 4D theory on
the brane. This conclusion has been reached by Binetruy \textit{et al.} \cite{Langlois} 
by examining the solutions to Einstein
equation in five dimensions on an $S^{1}/Z_{2}$ orbifold, with
matter on two branes and no cosmological constant on bulk or
brane. This issue was resolved by Cs\'aki \textit{et al.} \cite{Csaki} 
using time independent stabilized radion field.

The RS model consists of a spacetime with $S^{1}/Z_{2}$ orbifold
symmetry, two branes with opposite tension reside on orbifold
fixed point such that the spacetime metric has a redshift factor
depending exponentially on radius of the compactified dimension.
The lowering of Planck scale on negative tension brane also
depends on this compactified radius exponentially. For
$kr_{c}=12$, where $r_{c}$ is the compactification radius and $k$
is the parameter of same order as Planck scale $M$, the weak scale
is dynamically generated from this fundamental scale. The
Kaluza-Klein excitations have TeV mass splitting and couplings \cite{Randall2}. 
In this scenario (as presented in
\cite{Randall1}), $r_{c}$ is associated with vacuum expectation
value of a massless four dimensional scalar field. Since modulus
field has zero potential, the stabilization of $r_{c}$ cannot be
determined from the dynamics of the model. Subsequently Goldberger
and Wise (GW) \cite{Wise2} provided a stabilization
mechanism by introducing a bulk scalar field to generate a
potential for the modulus. However it does not address the
dynamics of this stabilization mechanism. Recently many other variants of the 
stabilization mechanisms are presented 
\cite{Binetruy,Cline1,Patil,Nasri,Carroll}. 
The stabilization mechanism has also been discussed from the point of view 
of gauge theories, string inspired models in many other works 
\cite{Kar,Cline2,Alberghi,Coley,Berndsen}.

In this work we shall first consider the Einstein equation with a metric ansatz having 
FRW structure in 4D sector and time dependent
radion field with energy density on both the branes. From Einstein's equation we first 
determine the averaged Einstein equations and then the equation for the radion field to 
find the dynamics of the radion field at early universe. In this context 
we have shown the equivalence of our result with the covariant 
curvature approach adopted in Ref. \cite{Soda} and \cite{Shiromizu}. 
It turns out that this 
evolution of radion field depends on the energy density on the visible brane. 
We have then obtained the evolution of the scale factor which depends on 
time evolution of the radion field. It has been shown that scale factor 
has an inflationary solution, which comes from the compactification of 
the extra dimension to a small value.
Finally we generalize the GW stabilization mechanism to time dependent 
radion field and have found the stabilized value of the radion field, 
which coincides with Goldberger-Wise solution.
The paper ends with a short discussion on our results.

%%%%%%%%%%%%%%%%%%%%%%%%%%%%%%%%%%%%%%%%%%%%%
%%%%%%%%%%%%%%%%%%%%%%%%%%%%%%%%%%%%%%%%%%%%%
\section{Einstein equations and solution}\label{braneeinst}

In this section we shall discuss the Einstein equation in RS model
with time dependent radion field along with matter on 
brane and a bulk cosmological constant. Using Einstein equation
and following the procedure used in the work of Cs\'aki \textit{et al.} 
\cite{Csaki} we find four-dimensional Friedman-like equations induced on
both branes and try to find radion time dependence.
The action for our system is given by,
%%%%%%%%%%%%%%%%%%%%%%%%%%%%%%%%%%%%%%%%%%%%%%
\begin{eqnarray}\label{braneg1a}
S&=&\int d^{5}x \sqrt{-\tilde{g}}\left(2M^{3}R-\Lambda \right)
\nonumber
\\
&+&\int d^{4}x \sqrt{-g_{hid}}~\left\lbrace \mathcal{L}_{hid}-V_{hid}\right\rbrace
\nonumber
\\
&+&\int d^{4}x \sqrt{-g_{vis}}~\left\lbrace \mathcal{L}_{vis}-V_{vis}\right\rbrace
\end{eqnarray}
%%%%%%%%%%%%%%%%%%%%%%%%%%%%%%%%%%%%%%%%%%%%%
where the first term corresponds to bulk part of the action with a bulk cosmological constant. 
The second part represents action on the hidden or Planck brane with $V_{hid}$ as the 
brane tension and $\mathcal{L}_{hid}$ as the matter Lagrangian on the hidden 3-brane. 
Similar considerations holds for the third term as well for visible or TeV brane. 
The quantity $M$ corresponds to five dimensional Planck scale.

The metric ansatz for the five-dimensional space time is taken as,
%%%%%%%%%%%%%%%%%%%%%%%%%%%%%%%%%%%%%
\begin{eqnarray}\label{b1}
ds^{2}&=&e^{-2A(\phi ,t)}\left[-dt^{2}+a^{2}(\phi ,t)\left(dx^{2}+dy^{2}+dz^{2}\right)\right]+r^{2}(\phi ,t)d\phi ^{2}
\nonumber
\\
&\equiv & \tilde{g}_{AB}(x,\phi)dx^{A}dx^{B}
\end{eqnarray}
%%%%%%%%%%%%%%%%%%%%%%%%%%%%%%%%%%%%%%%
The two branes are being located at $\phi =0$ and $\phi =\pi$. 
The Einstein tensor for this metric is given by component-wise,
%%%%%%%%%%%%%%%%%%%%%%%%%%%%%%%%%%%%%
\begin{eqnarray}
G_{00}&=&3\left( \dot{A}^{2}-\dot{A}\frac{\dot{r}}{r} \right)+
3\left( \frac{\dot{a}^{2}}{a^{2}}+\frac{\dot{a}}{a} 
\frac{\dot{r}}{r}- 2 \dot{A} \frac{\dot{a}}{a} \right)
\nonumber
\\
&+&3\frac{e^{-2A}}{r^{2}}\left( A^{''}-2(A^{'})^{2}-A^{'} 
\frac{r^{'}}{r}+ 4A^{'} \frac{a^{'}}{a}-\frac{a^{''}}{a}-
(\frac{a^{'}}{a})^{2}+\frac{a^{'}}{a}\frac{r^{'}}{r} \right);
\label{b2a}
\nonumber
\\
G_{ii}&=&a^{2}\left(2\ddot{A}-\dot{A}^{2}+\dot{A}\frac{\dot{r}}{r} \right)+
a^{2}\left( -2\frac{\ddot{a}}{a}-\frac{\dot{a}^{2}}{a^{2}}+
4\dot{A}\frac{\dot{a}}{a}-\frac{\ddot{r}}{r}-2\frac{\dot{a}}{a}\frac{\dot{r}}{r}\right)
\nonumber
\\
&+&a^{2} \frac{e^{-2A}}{r^{2}}\left[ 3\left( -A^{''}+2(A^{'})^{2}+
A^{'}\frac{r^{'}}{r}\right)+\left( 2\frac{a^{''}}{a}-8A^{'}\frac{a^{'}}{a}+
(\frac{a^{'}}{a})^{2}-2\frac{a^{'}}{a}\frac{r^{'}}{r} \right) \right];
\label{b2b}
\nonumber
\\
G_{04}&=&3\dot{A}^{'}-3A^{'}\frac{\dot{r}}{r}+3 \frac{a^{'}}{a}(\dot{A}+\frac{\dot{r}}{r})-3 \frac{\dot{a}^{'}}{a};
\label{b2c}
\nonumber
\\
G_{44}&=&6(A^{'})^{2}+3\left[\left( \frac{a^{'}}{a}\right)^{2}-3\frac{a^{'}}{a}A^{'}\right]+
r^{2}e^{2A}\left( 3\ddot{A}-3\dot{A}^{2}-3\frac{\ddot{a}}{a}-3\frac{\dot{a}^{2}}{a^{2}}+9\dot{A}\frac{\dot{a}}{a}\right)
\label{b2d}
\end{eqnarray}
%%%%%%%%%%%%%%%%%%%%%%%%%%%%%%%%%%%%%
Here primes (overdots) denote derivatives with respect to $\phi (t)$. 
The contribution to energy momentum tensor from bulk cosmological constant has the form,
%%%%%%%%%%%%%%%%%%%%%%%%%%%%%%%%%%%%% 
\begin{eqnarray}\label{b3}
T_{ab}^{bulk}=\tilde{g}_{ab}\Lambda
\end{eqnarray}
%%%%%%%%%%%%%%%%%%%%%%%%%%%%%%%%%%%%%
and for the branes we readily obtain,
%%%%%%%%%%%%%%%%%%%%%%%%%%%%%%%%%%%
\begin{eqnarray}\label{b4}
T_{a}^{b,~brane}&=&\frac{\delta (\phi)}{r}diag\left(V_{hid}+\rho _{hid},V_{hid}-p_{hid},V_{hid}-p_{hid},V_{hid}-p_{hid},0\right)
\nonumber
\\
&+&\frac{\delta (\phi -\pi)}{r}diag\left(V_{vis}+\rho _{vis},V_{vis}-p_{vis},V_{vis}-p_{vis},V_{vis}-p_{vis},0\right)
\end{eqnarray}
%%%%%%%%%%%%%%%%%%%%%%%%%%%%%%%%%%%
where $'$ denotes derivative with respect to $\phi$ and overdot
denotes derivative with respect to $t$. Here $\Lambda$ is the bulk
cosmological constant and $V_{vis}$ and $V_{hid}$ are the constant
"vacuum energy" on the 3-branes which acts as gravitational source
even in absence of particle excitations. The parameters $V_{hid}$,
$V_{vis}$ and $\lambda$ are related to a single scale k such that,
%%%%%%%%%%%%%%%%%%%%%%%%%%%%%%%%%%%%%%%%
\begin{eqnarray}\label{b6}
\left.\begin{array}{c}
V_{hid}=-V_{vis}=24M^{3}k\\

\Lambda =-24M^{3}k^{2}
\end{array} \right \}
\end{eqnarray}
%%%%%%%%%%%%%%%%%%%%%%%%%%%%%%%%%%%%%%%%%%%%%
Now the quantities $\rho _{hid}$ and $p_{hid}$ are the density and pressure 
of matter on positive tension brane. Also the quantities $\rho _{vis}$ and 
$p_{vis}$ are respective quantities in TeV brane. In the limit 
$\rho _{hid},p_{hid},\rho _{vis},p_{vis}\rightarrow 0$ we should recover 
the static Randall-Sundrum solution. However initial investigation of 
cosmology of brane world models shows some inconsistency since there 
appears a constraint on the bare matter density in Planck and TeV brane. 
Also from Hubble parameter it appears that $\rho _{hid}>0$ which in turn 
imply that energy density on TeV brane is negative. These results were 
discussed in \cite{Csaki} and physical interpretations were given in a 
quiet general sense. 
From the above Einstein equations it is quiet clear that if we assume that 
warp factor $A$ depends only on the extra space-time co-ordinate $\phi$, 
the scale factor and the radion field depend on $t$.
Then from the $G_{04}$ equation one finds $\dot{r}=0$. Hence the 
radion field cannot have a time dependence and therefore cannot evolve 
dynamically. Also from the other equations we see that the scale 
factor becomes time independent leading to the static Randall-Sundrum solution.

%%%%%%%%%%%%%%%%%%%%%%%%%%%%%%%%%%%%%%%%%%%
\subsection{Averaged Einstein equations}\label{braneavg}
%%%%%%%%%%%%%%%%%%%%%%%%%%%%%%%%%%%%%%%%%%%%%
In this subsection we present averaged Einstein equations over the bulk to demonstrate that 
without a stabilizing potential the system become over constrained if we require the 
modulus to be static. We apply the following expansion around RS solution:
%%%%%%%%%%%%%%%%%%%%%%%%%%%%%
\begin{eqnarray}\label{b7a}
a(\phi ,t)&=&a(t)\left[1+\delta a(\phi ,t)\right]
\nonumber
\\
r(\phi ,t)&=&r(t)\left[1+\delta r(\phi ,t)\right]
\nonumber
\\
A(\phi ,t)&=&kr(t)\mid \phi \mid \left[1+\delta A(\phi ,t)\right]
\end{eqnarray}
%%%%%%%%%%%%%%%%%%%%%%%%%%%%%
The warp factor can be given by introducing the $A(\phi ,t)$ as,
%%%%%%%%%%%%%%%%%%%%%%%%%%
\begin{equation}\label{b7b}
\Omega \equiv \Omega (\phi ,r(t))=e^{-kr(t)\mid \phi \mid}; ~~~\Omega _{r}\equiv e^{-kr(t)\pi}
\end{equation}
%%%%%%%%%%%%%%%%%%%%%%%%%%%%%
The value of $\Omega _{r}$, when $r=r_{c}=\textrm{constant}$ would be given by $\Omega _{c}$. 
All the quantities $\delta a$, $\delta A$ and $\delta r$ 
are perturbations on the RS background along with $\rho _{hid}$ and $\rho _{vis}$. 
Therefore in the above expansion we can make the assumption that $\delta a$, $\delta A$ and $\delta r$ 
are linear functions of energy densities $\rho _{hid}$ and $\rho _{vis}$ only \cite{Csaki}. 
The linearity assumption is necessary from the requirement 
that in the limit $\rho _{hid} \rightarrow 0$ and $\rho _{vis} \rightarrow 0$ one 
should recover RS solution as well as when all the 
metric perturbations $\delta a, \delta A$ and $\delta r$ are set to zero. 
This implies that an expansion in the brane matter energy density is equivalent to 
an expansion in metric perturbations. This amounts to assume $\delta a$, $\delta A$ and $\delta r$ 
to be linear in the matter energy density as both are perturbations on 
the RS background. With implication that time derivatives of the 
perturbations are higher order in energy density, which, then can be neglected. 
However we shall work with all possible orders in $r(t)$ and its time derivative 
in this subsection.

For completeness we also include a radion potential following the work by 
Cs\'aki \textit{et al.} \cite{Csaki}:
%%%%%%%%%%%%%%%%%%%%%%%%%%%%%%%%%
\begin{equation}\label{b7c}
a^{3}V_{r}(r)\equiv -r(t)\int d\phi \Omega ^{4}\mathcal{L_{R}}
\end{equation}
%%%%%%%%%%%%%%%%%%%%%%%%%%%%%%%%%
in the following computations and shall set $V_{r}=0$ whenever desired. Here 
the quantity $\mathcal{L_{R}}$, the potential Lagrangian is generated from 
some dynamics (may be scalar field, see section \ref{branerad} for a detailed discussion).

For classical electromagnetism on a manifold without any boundary we readily 
obtain by integrating $\nabla . E=\rho$ that total charge must vanish. Here 
also to find some topological constraint, we use analog of Gauss' law in 
Einstein theory. For example consider the following integral \cite{Csaki},
%%%%%%%%%%%%%%%%%%%%%%%%%%%%%%%%%%%%%%%
\begin{equation}\label{b12d}
\int d\phi \Omega ^{4}G_{0}^{0}=\frac{1}{4M^{3}}\int d\phi \Omega ^{4} T_{0}^{0}
\end{equation}
%%%%%%%%%%%%%%%%%%%%%%%%%%%%%%%%%%%%%%%%%%
However this equation does not imply a topological constraint, rather is 
combined with other averaged equations to give the constrained energy densities. 
This actually follows from requiring a static extra dimension without radion 
potential. Note that this implies that radion field could evolve even without 
any stabilizing potential.

Then substitution of the above expansion, equation (\ref{b7a}) into equation 
(\ref{b12d}) and integrating we arrive at,
%%%%%%%%%%%%%%%%%%%%%%%%%%%%%%%%%%%
\begin{equation}\label{b7d}
\left(\frac{\dot{a}}{a}\right)^{2}+2kr_{c}\pi \frac{\Omega _{r}^{2}}{1-\Omega _{r}^{2}}
\frac{\dot{a}}{a}\frac{\dot{r}}{r}-k^{2}r^{2}\pi ^{2}
\frac{\Omega _{r}^{2}}{1-\Omega _{r}^{2}}\left(\frac{\dot{r}}{r}\right)^{2}=
\frac{k}{12M^{3}}\frac{1}{1-\Omega _{r}^{2}}
\left[\rho _{hid}+\rho _{vis}\Omega _{r}^{4}+V_{r}(r)\right]+\mathcal{O}(\epsilon ^{2})
\end{equation}
%%%%%%%%%%%%%%%%%%%%%%%%%%%%%%%%%%%%%
where $\epsilon ^{2}= \mathcal{O}(\delta a^{2}, \delta A^{2}, \delta r^{2})$. 
There are no correction to the Hubble parameter squared which is linear 
in the perturbations. Also the above equation reduces to conventional 
FRW solutions when the energy density in the radion is small. In this 
limit the behavior of scale factor is determined by the quantity 
$\rho _{hid}+\rho _{vis}\Omega _{r}^{4}$, i.e. it depends on both 
energy density in Planck and TeV brane. Hence for small oscillation 
of the radion field the expansion of the universe is determined by 
ordinary FRW cosmology. Hence the bulk averaged $G_{00}$ equation 
results in ordinary Hubble law.
We can repeat the above procedure for the spatial $G_{ij}$ components 
as well which resulted in the following equation:
%%%%%%%%%%%%%%%%%%%%%%%%%%%%%%%%%%%
\begin{eqnarray}\label{b7e}
2\left(\frac{\ddot{a}}{a}\right)&+&\left(\frac{\dot{a}}{a}\right)^{2}+
4k\pi r\frac{\Omega _{r}^{2}}{1-\Omega _{r}^{2}}
\frac{\dot{a}}{a}\frac{\dot{r}}{r}-k^{2}r^{2}\pi ^{2}
\frac{\Omega _{r}^{2}}{1-\Omega _{r}^{2}}\left(\frac{\dot{r}}{r}\right)^{2}
\nonumber
\\
&+&2kr\pi \frac{\Omega _{r}^{2}}{1-\Omega _{r}^{2}} 
\frac{\ddot{r}}{r}=-\frac{k}{4M^{3}}\frac{1}{1-\Omega _{r}^{2}}
\left[p_{hid}+p_{vis}\Omega _{r}^{4}-V_{r}(r)\right]+\mathcal{O}(\epsilon ^{2})
\end{eqnarray}
%%%%%%%%%%%%%%%%%%%%%%%%%%%%%%%%%%%%
Note that in the limit when radion field is static and have no associated 
radion potential, the above equation reduces to standard pressure equation 
in FRW cosmology. This is also in accord with the averaged $G_{00}$ equation 
due to the fact that there exists no correction of $\mathcal{O}(\epsilon)$.

However for the unaveraged linearized $G_{44}$ equation the perturbations 
$\mathcal{O}(\epsilon)$ appear. Then if we follow the respective jump 
conditions as given in \cite{Langlois} then at $\phi =0$, the following 
equation results \cite{Csaki},
%%%%%%%%%%%%%%%%%%%%%%%%%%%%%%%%%%%
\begin{equation}\label{b7f}
-\frac{k^{2}}{24M^{3}}\left(-\rho _{hid}+3p_{hid}\right)-
\left(\frac{\dot{a}^{2}}{a^{2}}+\frac{\ddot{a}}{a}\right)=
\frac{1}{12M^{3}}T_{5}^{5}\mid _{0}
\end{equation}
%%%%%%%%%%%%%%%%%%%%%%%%%%%%%%%%%%%%
In the absence of a radion potential however the system is 
over constrained provided a static radion field solution is imposed. 
To observe this more closely we can eliminate the scale factor using 
equations (\ref{b7d}) and (\ref{b7e}) which ultimately leads to a static solution of r,
%%%%%%%%%%%%%%%%%%%%%%%%%%%%%%
\begin{equation}\label{b7g}
\left(-3p_{vis}+\rho _{vis} \right)\Omega _{r}^{2}=\left(3p_{hid}-\rho _{hid}\right)
\end{equation}
%%%%%%%%%%%%%%%%%%%%%%%%%%%%%%%%
This along with conservation of energy leads to a further fine tuning and the above constraint leads to,
%%%%%%%%%%%%%%%%%%%%%%%%%%%%%%%%
\begin{equation}\label{b7h}
\rho _{hid}=-\Omega _{r}^{2}\rho _{vis}
\end{equation}
%%%%%%%%%%%%%%%%%%%%%%%%%%%%%%%%%%%
This constraint arises due to requiring $r=constant$ without a radion potential. 
With the above assumptions system becomes over constrained and fine tuning of 
energy densities is necessary to maintain a static solution in bulk.
%%%%%%%%%%%%%%%%%%%%%%%%%%%%%%%%%%%%%%%%%%%%%%%%%%%%%%%%%%%%%%%%%%%%%%%%%%%%%%%%%%%%%%%%%%%%%
%%%%%%%%%%%%%%%%%%%%%%%%%%%%%%%%%%%%%%%%%%%%%%%%%%%%%%%%%%%%%%%%%%%%%%%%%%%%%%%%
\subsection{Equivalence with covariant curvature formalism}\label{branecov}

The linearized theory in brane world models were discussed in \cite{Garriga,Rubakov}. In Ref. \cite{Soda} 
effective Einstein equation was derived at low energy scale. 
However that derivation was a metric based approach. 
Here we will follow the covariant curvature approach illustrated in 
Ref. \cite{Shiromizu}. This covariant curvature formalism 
gives an effective gravitational equation on the branes. 
We will use the metric ansatz as given by equation (\ref{b1}), 
in the following form,
%%%%%%%%%%%%%%%%%%%%%%%%%%%%%%%%%%%%%%%%%%%
\begin{equation}\label{bc1}
ds^{2}=g_{\mu \nu}dx^{\mu}dx^{\nu}+r^{2}d\phi ^{2}
\end{equation}
%%%%%%%%%%%%%%%%%%%%%%%%%%%%%%%%%%%%%%%%%%%%
Then the proper distance between the brane is given by 
$d_{0}=\pi r(t)$ and $g_{\mu \nu}$ is the induced metric 
over $\phi =$ constant hypersurfaces.
Then following the procedure as presented by in 
Ref. \cite{Shiromizu} and \cite{Maeda}) 
we obtain the Gauss-Codazzi equations as,
%%%%%%%%%%%%%%%%%%%%%%%%%%%%%%%%%%%%%%%%%%%%%%%%%
\begin{eqnarray}\label{bc2}
^{(4)}G_{\mu \nu}&=&3k^{2}\delta _{\nu}^{\mu}+KK^{\mu}_{\nu}-K^{\mu}_{\alpha}K^{\alpha}_{\nu}
\nonumber
\\
&-& \frac{1}{2}\delta^{\mu}_{\nu}\left(K^{2}-K^{\alpha}_{\beta}K^{\beta}_{\alpha}\right)-E^{\mu}_{\nu}
\end{eqnarray}
%%%%%%%%%%%%%%%%%%%%%%%%%%%%%%%%%%%%%%%%%%%%%%%%%
and
%%%%%%%%%%%%%%%%%%%%%%%%%%%%%%%%%%%
\begin{equation}\label{bc3}
 D_{\mu}K^{\mu}_{\nu}-D_{\mu}K=0
\end{equation}
%%%%%%%%%%%%%%%%%%%%%%%%%%%%%%%%%
with $D_{\mu}$ being the covariant derivative on the $\phi =$ constant hypersurfaces. 
Here $k$ is the bulk curvature, $^{(4)}G_{\mu \nu}$ is the 4-dimensional Einstein tensor 
and $K_{\mu \nu}$ is the extrinsic curvature of the hypersurfaces defined as,
%%%%%%%%%%%%%%%%%%%%%%%%%%%%%%%%%%%
\begin{equation}\label{bc4}
 K_{\mu \nu}=\frac{1}{2}\pounds _{n}q_{\mu \nu}=\nabla _{\mu}n_{\nu}+n_{\mu}a_{\nu}
\end{equation}
%%%%%%%%%%%%%%%%%%%%%%%%%%%%%%%%%%%%
where we have $n=r^{-1}\partial _{\phi}$ and $a^{\mu}=n^{\nu}\nabla _{\nu}n^{\mu}$. 
$E_{\mu \nu}$ is the projected Weyl tensor defined by, 
$E^{\mu}_{\nu}=^{(5)}C_{\mu \alpha \nu \beta}n^{\alpha}n^{\beta}$, 
with $^{(5)}C_{\mu \alpha \nu \beta}$ is the 5-dimensional Weyl tensor. 
The respective jump conditions are given by,
%%%%%%%%%%%%%%%%%%%%%%%%%%%%%%%%%%%%%%%%%
\begin{eqnarray}\label{bc5}
\left[K^{\mu}_{\nu}-\delta ^{\mu}_{\nu}K \right]_{\phi=0}
&=&-\frac{1}{8M^{3}}\left(-V_{hid}\delta ^{\mu} _{\nu}+T_{1~\nu}^{\mu} \right)
\\
\left[K^{\mu}_{\nu}-\delta ^{\mu}_{\nu}K \right]_{\phi=\pi}
&=&-\frac{1}{8M^{3}}\left(-V_{vis}\delta ^{\mu} _{\nu}+T_{2~\nu}^{\mu} \right)
\end{eqnarray}
%%%%%%%%%%%%%%%%%%%%%%%%%%%%%%%%%%%%%%%%%
where $T_{1~\nu}^{\mu}$ and $T_{2~\nu}^{\mu}$ are the energy momentum tensor of the 
hidden and visible brane respectively. $V_{hid}$ and $V_{vis}$ are the 
brane tensions of hidden and visible branes 
respectively.
Then following the procedure adopted in Ref. \cite{Soda} and \cite{Shiromizu} we arrive at the 
low-energy effective theory with effective Einstein equation on visible brane,
%%%%%%%%%%%%%%%%%%%%%%%%%%%%%%%%%%%%%%%%%%%
\begin{eqnarray}\label{bc6}
^{(4)}G^{\mu}_{\nu}&=&\frac{k}{4M^{3}}\frac{1}{\Phi}T_{2~\nu}^{\mu}
+\frac{k}{4M^{3}}\frac{(1+\Phi)^{2}}{\Phi}T_{1~\nu}^{\mu}
\nonumber
\\
&+&\frac{1}{\Phi}\left(D^{\mu}D_{\nu}-\delta ^{\mu}_{\nu}D^{2}\Phi \right)
\nonumber
\\
&+&\frac{\omega(\Phi)}{\Phi^{2}}\left(D^{\mu}\Phi D_{\nu}\Phi
-\frac{1}{2}\delta ^{\mu}_{\nu}\left(D\Phi \right)^{2} \right)
\end{eqnarray}
%%%%%%%%%%%%%%%%%%%%%%%%%%%%%%%%%%%%%%%%%%%%%%
where $\Phi=e^{2kr(t)\pi}-1=e^{2A}-1$ and $\omega (\Phi)=-\frac{3}{2}\frac{\Phi}{1+\Phi}$. 
This is the effective equation in leading order of brane to bulk curvature ratio. 
Then using the metric ansatz given by equation (\ref{b1}) we readily obtain 
the time-time component of the above equation to yield,
%%%%%%%%%%%%%%%%%%%%%%%%%%%%%%%%%%%%%%%%%%%%%%%
\begin{eqnarray}\label{bc7}
\Omega ^{-2}\Big[3\left(\frac{\dot{a}}{a}\right)^{2}&-&6k\pi r \frac{\dot{a}}{a}\frac{\dot{r}}{r}+
3k^{2}\pi ^{2} r^{2}\left(\frac{\dot{r}}{r} \right)^{2}\Big]
=\frac{k}{4M^{3}}\frac{1}{\Omega ^{-2}-1}\left(\rho _{vis}+\Omega ^{-4}\rho _{hid}\right)
\nonumber
\\
&+&6\frac{\Omega ^{-4}}{\Omega^{-2}-1}k^{2}\pi ^{2}r^{2}\left(\frac{\dot{r}}{r} \right)^{2}
-6\frac{\Omega ^{-4}}{\Omega ^{-2}-1}k\pi r \frac{\dot{a}}{a}\frac{\dot{r}}{r}
-3\frac{\Omega ^{-4}}{\Omega^{-2}-1}k^{2}\pi ^{2}r^{2}\left(\frac{\dot{r}}{r} \right)^{2}
\end{eqnarray}
%%%%%%%%%%%%%%%%%%%%%%%%%%%%%%%%%%%%%%%%%%%%%%%%%%%%%%%%%%%%%%%%%%
which can be further simplified to yield,
%%%%%%%%%%%%%%%%%%%%%%%%%%%%%%%%%%%%%%%%%%%%%%%%%%%%%%%%%%%%%%%%%%
\begin{eqnarray}\label{bc8}
\left(\frac{\dot{a}}{a}\right)^{2}&+&2k\pi r\left(-1+\frac{1}{1-\Omega ^{2}}\right) 
\frac{\dot{a}}{a}\frac{\dot{r}}{r}-
k^{2}\pi ^{2} r^{2}\left(-1+\frac{1}{1-\Omega ^{2}}\right)\left(\frac{\dot{r}}{r} \right)^{2}
\nonumber
\\
&=& \frac{k}{12M^{3}}\frac{1}{1-\Omega ^{2}}\left(\rho _{hid}+\Omega ^{-4}\rho _{vis}\right)
\end{eqnarray}
%%%%%%%%%%%%%%%%%%%%%%%%%%%%%%%%%%%%%%%%%%%%%%%%%%%%%%%%%%%%%%%%%
which is precisely the time-time component of the averaged Einstein equation given by 
equation (\ref{b7d}). The quantity $\Omega$ used in 
the above equation is given by $\Omega = e^{-2k\pi r}$. 
Next we consider the space-space component of the effective equation on negative tension brane,
%%%%%%%%%%%%%%%%%%%%%%%%%%%%%%%%%%%%%%%%%%%%%%%%%%%%%%%%%%%%%%%%%%%%%
\begin{eqnarray}\label{bc9}
\Omega ^{-2}\Big[2\left(\frac{\ddot{a}}{a} \right)&+&\left(\frac{\dot{a}}{a}\right)^{2}
-4k\pi r\frac{\dot{a}}{a}\frac{\dot{r}}{r}-
k^{2}r^{2}\pi ^{2}\left(\frac{\dot{r}}{r}\right)^{2}-2k\pi r \frac{\ddot{r}}{r}\Big]
\nonumber
\\
&=&-\frac{k}{4M^{3}}\frac{1}{\Omega ^{-2}-1}\left(p _{vis}+\Omega ^{-4}p _{hid}\right)
-4k\pi r \frac{\Omega ^{-4}}{\Omega ^{-2}-1}
\frac{\dot{a}}{a}\frac{\dot{r}}{r}
\nonumber
\\
&-&k^{2}r^{2}\pi ^{2}\frac{\Omega ^{-4}}{\Omega ^{-2}-1}\left(\frac{\dot{r}}{r}\right)^{2}-
2k\pi r \frac{\Omega ^{-4}}{\Omega ^{-2}-1} \frac{\ddot{r}}{r}
\end{eqnarray}
%%%%%%%%%%%%%%%%%%%%%%%%%%%%%%%%%%%%%%%%%%%%%%%%%%%%%%%%%%%%%%%%%%%%%%%%%%%%%%
which can be simplified to yield the space-space part of 
averaged Einstein equations given by equation (\ref{b7e}). 
The same procedure can now be applied for the hidden brane to 
retrieve the same equations as (\ref{b7d}) and 
(\ref{b7e}). This proves the equivalence of the covariant curvature formalism with the averaged Einstein 
equation procedure. It must be stressed that this equivalence holds only when higher order 
terms can be neglected. For example, in the case of strong time dependence the higher order corrections 
in terms of brane to bulk curvature ratio cannot be neglected. 
In that situation the above equivalence may not be valid.

%%%%%%%%%%%%%%%%%%%%%%%%%%%%%%%%%%%%%%%%%%%%%%%%%%%%%%%%%%%%%%%%%%%%%%%%%%%%%%%%%%%%%%%%%%%%%%%%%555
%%%%%%%%%%%%%%%%%%%%%%%%%%%%%%%%%%%%%%%%%%%%%%%%%%%%%%%%%%%%%%%%%%%%%%%%
\subsection{Dynamics of the radion field}\label{branedyn}

To address the dynamics of compactification we first consider radion field to be time dependent 
and subsequently shall provide a stabilization mechanism. We shall use 
the averaged Einstein equations as given by equations
(\ref{b7d}) and (\ref{b7e}) and obtain averaged equations in terms of the
unknown functions $a(t)$ and $r(t)$ in terms of the energy densities in 
visible and Planck brane. For this purpose we use the equation of 
state $\rho _{i}=-p_{i}$. However in all the bulk
Einstein equations the right hand side gets canceled by the
respective terms on the left hand side, thus we need not to bother
about the boundary terms anymore. The different components of averaged
Einstein equation leads to:
%%%%%%%%%%%%%%%%%%%%%%%%%%%%%%%%%%%%%%%%%%
\begin{eqnarray}
\left(\frac{\dot{a}}{a}\right)^{2}&+&2kr\pi \frac{\Omega _{c}^{2}}{1-\Omega _{c}^{2}}
\frac{\dot{a}}{a}\frac{\dot{r}}{r}-k^{2}r^{2}\pi ^{2}\frac{\Omega _{c}^{2}}{1-\Omega _{c}^{2}}
\left(\frac{\dot{r}}{r}\right)^{2}=-\frac{k}{12M^{3}}\Omega _{c}^{2}\rho _{vis}
\label{b12a}
\\
2\left(\frac{\ddot{a}}{a}\right)&+&\left(\frac{\dot{a}}{a}\right)^{2}
+4k\pi r\frac{\Omega _{c}^{2}}{1-\Omega _{c}^{2}}
\frac{\dot{a}}{a}\frac{\dot{r}}{r}-k^{2}r^{2}\pi ^{2}\frac{\Omega _{c}^{2}}
{1-\Omega _{c}^{2}}\left(\frac{\dot{r}}{r}\right)^{2}+
2kr\pi \frac{\Omega _{c}^{2}}{1-\Omega _{c}^{2}} \frac{\ddot{r}}{r}
=-\frac{k}{4M^{3}}\Omega _{c}^{2}\rho _{vis}
\label{b12b}
\end{eqnarray}
%%%%%%%%%%%%%%%%%%%%%%%%%%%%%%%%%%%%%%%%%%%%
where we have taken the energy density and pressure such that, 
$\rho _{hid}=-\Omega _{c}^{2}\rho _{vis}$ and 
$p_{hid}=-\rho _{hid}$ along with $p_{vis}=-\rho _{vis}$.
However for the $44$ component we readily obtain that the unaveraged Einstein equation 
leads to an identity following the work by Cs\'aki \textit{et al.} \cite{Csaki}. 
Now equation (\ref{b12a}) can be solved for $\dot{a}/a$ such that, 
%%%%%%%%%%%%%%%%%%%%%%%%%%%%%%%%%%%%%%%%%%%%%%%%%%%%%%%%%%%%%%%%%%
\begin{equation}\label{b13}
\dot{a}/a=-k\pi r\frac{\Omega _{c} ^{2}}{1-\Omega _{c} ^{2}}\dot{r}/r\pm 
\sqrt{k^{2}\pi ^{2}r^{2}\frac{\Omega _{c} ^{2}}{\left(1-\Omega _{c}^{2}\right)^{2}}
\left(\dot{r}/r\right)^{2}-k/12M^{3}\Omega _{c}^{2}\rho _{vis}}
\end{equation}
%%%%%%%%%%%%%%%%%%%%%%%%%%%%%%%%%%%%%%%%%%%%%%%%%%
with the choice of small time scale such that,
%%%%%%%%%%%%%%%%%%%%%%%%%%%%%%%%%%%%%%%%%%%%%%%%
\begin{equation}\label{b14}
t\ll t_{valid}=\frac{90}{\Omega _{c}^{2}}\sqrt{\frac{12M^{3}}{k\rho _{vis}}}
\frac{\left(1-\Omega _{c}^{2}\right)}{\left(1+\Omega _{c}^{2}\right)}
\end{equation}
%%%%%%%%%%%%%%%%%%%%%%%%%%%%%%%%%%%%%%%%%%%%%%%%
we can readily observe that the only physical parameter present in the above expression is 
$\left(\rho _{vis}\frac{k}{M^{3}} \right)^{-1}$. Now from equation (\ref{b6}) we readily observe 
that this actually imply that $\frac{\rho _{vis}V_{vis}}{M^{6}}$ 
should be large. Note that brane tension is order 
of $M^{3}$, thus for this to valid $\rho _{vis}$ should be greater than $M^{3}$, 
which is true only before the Planck time, and that is precisely when the inflation occurs. 
This justifies our use of small time approximation and presenting this solution as 
an inflationary scenario. Also in the above expression $t_{valid}$ represents 
the time scale upto which the small time approximation would remain valid.
Then under this small time scale we can make a linear choice for the 
time evolution of the radion field and that yield the connection between 
the matter energy density and the radion field such that,
%%%%%%%%%%%%%%%%%%%%%%%%%%%%%%%%%%%%%%%
\begin{equation}\label{bm3b}
\dot{r}=-\sqrt{\frac{\rho _{vis}}{12M^{3}k\pi ^{2}}}\left(1-\Omega _{c}^{2}\right)
\end{equation}
%%%%%%%%%%%%%%%%%%%%%%%%%%%%%%%%%%%%%%%%%%
It is now evident from the above equation that if $\rho _{vis}$ is zero i.e. there is no energy 
density on the visible brane, the radion field cannot evolve with time. 
Thus the radion field would have been zero without Goldberger-Wise stabilization mechanism. 
The stabilized value $r=r_{c}$ can be obtained by introducing a bulk scalar field. In the case of 
time dependent situation such a stabilization mechanism has been presented in section \ref{branerad}. 

However we should also mention that the equation given by (\ref{bm3b}) 
only holds if the radion field evolves 
dynamically with time. After the radion field become stabilized the above equation would not be valid. 
Then we have to resort to the unaveraged Einstein equation and obtain 
the evolution of the scale factor from those equations directly.

Now we will consider the fact that $\rho _{vis}=\textrm{constant}$. 
The choice $\rho _{vis}=-p_{vis}=\textrm{constant}$ 
actually imply the standard dark energy candidate.
Then for the differential equation as presented in equation (\ref{bm3b}) we readily obtain,
%%%%%%%%%%%%%%%%%%%%%%%%%%%%%%%%%%%%%%
\begin{equation}\label{bm4}
r(t)=r_{0}-\sqrt{\frac{\rho _{vis}}{12M^{3}k\pi ^{2}}} \left(1-\Omega _{c}^{2}\right)t
\end{equation}
%%%%%%%%%%%%%%%%%%%%%%%%%%%%%%%%%
Now having obtained that solution we now proceed to determine the variation of the scale factor with time. 
For that purpose we use the averaged equation as presented by equations (\ref{b12a}) and (\ref{b13}) 
along with the radion field time evolution equation (\ref{bm3b}) leading to:
%%%%%%%%%%%%%%%%%%%%%%%%%%%%%%%%%%%%%
\begin{equation}\label{bm3a}
\frac{\dot{a}}{a}=-kr\pi\frac{\Omega _{c}^{2}}{1-\Omega _{c}^{2}}\frac{\dot{r}}{r}
\end{equation}
%%%%%%%%%%%%%%%%%%%%%%%%%%%%%%%%%%%%%%%%
The equations given by (\ref{b12a}) and (\ref{b12b}) leads 
finally to the fact that, radion field evolve by 
the energy density on the brane, which in turn make the scale factor to 
evolve. Having observed that $\dot{r}$ 
is negative which implies that radion field depletes with time, 
explaining the small value of extra dimension in the present epoch.
From the above equations we observe that when the radion field has
no time dependence then the usual Friedman equations are obtained with energy
density and pressure. In that limit our result matches with the
result of Cs\'aki \textit{et al.} \cite{Csaki}.

Then using the above solution the following solution for scale factor, 
we obtain from equation (\ref{bm3a}) which is,
%%%%%%%%%%%%%%%%%%%%%%%%%%%%%%
\begin{equation}\label{bm4a}
a(t)=a_{0}e^{H t};~~~~~H =\Omega _{c} ^{2}\sqrt{\frac{k\rho _{vis}}{12M^{3}}}
\end{equation}
%%%%%%%%%%%%%%%%%%%%%%%%%%%%%%
It is generally believed that the
universe has undergone a exponential increase of scale factor
which resulted in a very smooth universe at large scales solving
the flatness-oldness problem, due to quantum fluctuation. The
scale factor for that epoch is taken as, $a(t)\propto e^{\kappa t}$,
where $\kappa$ is a constant and denotes the inverse time in
which the universe has grown to $e$ times the previous value. 
Then the e-folding parameter is given by,
%%%%%%%%%%%%%%%%%%%%%%%%%%%%%%%%%%%%%%%%%%%%%
\begin{equation}\label{bm9}
N=\int _{t_{i}}^{t_{f}}Hdt=\Omega _{c} ^{2}\sqrt{\frac{k\rho _{vis}}{12M^{3}}}\Delta t
\end{equation}
%%%%%%%%%%%%%%%%%%%%%%%%%%%%%%%%%%%%%%%%%
Thus we have a situation where, as the brane gets inflated, the radion field decreases in 
magnitude making it consistent with recent observations. The observations suggest that e-folding parameter 
should be $\simeq 60$. Then by equating $N=60$ we observe that,
%%%%%%%%%%%%%%%%%%%%%%%%%%%%%%%%%%%%%%%%%%%%%%%
\begin{equation}\label{TimeRad}
\Delta t \simeq \frac{60}{\Omega _{c}^{2}}\sqrt{\frac{12M^{3}}{k\rho _{vis}}} \ll t_{valid}
\end{equation}
%%%%%%%%%%%%%%%%%%%%%%%%%%%%%%%%%%%%%%%%%%%%%%%%%%%%%%
Hence the use of small time approximation is justified and it remains valid 
during the whole period of inflation \cite{Alexander}. 
In the above inflationary scenario the radion field plays the roll 
of slow-roll parameter. As the radion field gradually tends to the 
stabilized value $r_{c}$, its time dependence 
is the primary cause of evolution of the scale factor. 
When the radion field gets stabilized, then $\dot{r}$ vanishes making 
scale factor time independent implying an exit from the inflationary phase. 
This time dependent stabilization is addressed 
in section \ref{branerad}.

Now we try for a solution of the form 
$r(t)=r_{0}-\sqrt{\frac{\rho _{vis}}{12M^{3}k\pi ^{2}}} \left(1-\Omega _{c}^{2}\right)t+\beta t^{2}$ 
in order that it should satisfy equation (\ref{b12b}). 
Since we are considering very early times in this discussion 
we readily obtain that while taking first derivative the term linear in time does not contribute. 
Hence using this ansatz in the above equation we readily obtain,
%%%%%%%%%%%%%%%%%%%%%%%%%%%%%%%%%%%%%
\begin{equation}\label{bm10}
r(t)=r_{0}-\sqrt{\frac{\rho _{vis}}{12M^{3}k\pi ^{2}}} \left(1-\Omega _{c}^{2}\right) t-
\left\lbrace \frac{\left(1-\Omega _{c}^{2}\right)\left(2-\Omega _{c}^{2}\right) 
\rho _{vis}}{48 M^{3}\pi}\right\rbrace t^{2}
\end{equation}
%%%%%%%%%%%%%%%%%%%%%%%%%%%%%%%%%%%%
From equation (\ref{bm10}) it might appear that it is not compatible with equation (\ref{bm4}), 
however we should note that equation (\ref{bm3b}) has been obtained on the ground that 
all terms that contain $t$ has been neglected which is justified for the small time 
scale we are working with (This is the motivation for introducing small 
time approximation in this work). Hence the second term in equation (\ref{bm10}) 
has no influence on the previous expression. 

Further to see that the coefficient of $t^{2}$ 
in Eq. (\ref{bm10}) should be sub-leading, 
we use time derivative of the radion field to get:
%%%%%%%%%%%%%%%%%%%%%%%%%%%%%%%%%%%%%%%%%%%%%%%%%%%%%%%%%%%%%%%%%%%%%%%%
\begin{equation}
\frac{\left(1-\Omega _{c}^{2}\right)\left(2-\Omega _{c}^{2}\right) 
\rho _{vis}}{48 M^{3}\pi}\times 2t \ll 
\sqrt{\frac{\rho _{vis}}{12M^{3}k\pi ^{2}}} \left(1-\Omega _{c}^{2}\right)
\end{equation}
%%%%%%%%%%%%%%%%%%%%%%%%%%%%%%%%%%%%%%%%%%%%%%%%%%%%%%%%%%%%%%%%%%%%%%%%
The above inequality can be simplified further 
using Eq. (\ref{TimeRad}) leading to: 
$t\ll \Omega _{c}^{2}t_{valid}$. Since $\Omega _{c}^{2}$ is already very small 
this translates into $t\ll t_{valid}$. Hence the small time approximation makes 
the $t^{2}$ term sub-leading.

%%%%%%%%%%%%%%%%%%%%%%%%%%%%%%%%%%%%%%%%%%%%%%%%%%%%%%%%%%%%%%%%%%%%%%%%%%%%%
\begin{figure}
\begin{center}

\includegraphics[height=2in, width=3in]{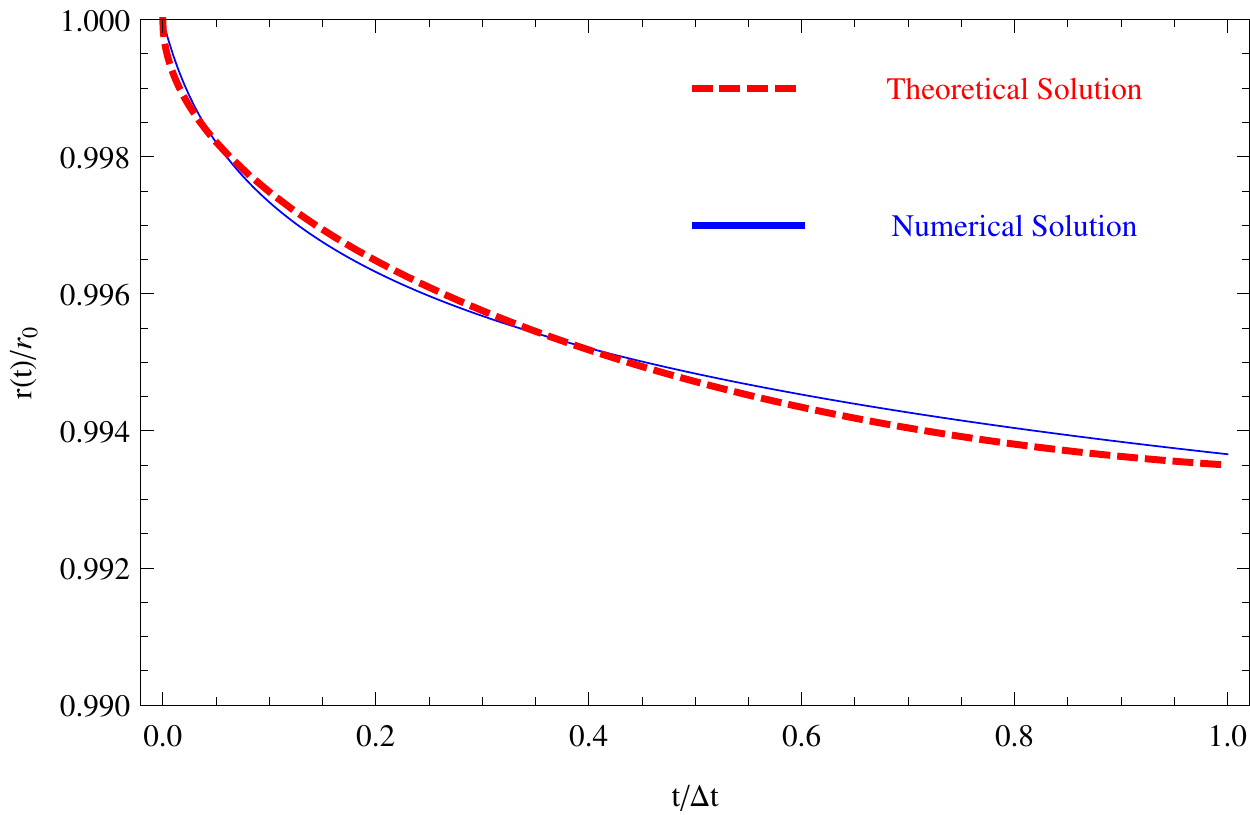}~~
\includegraphics[height=2in, width=3in]{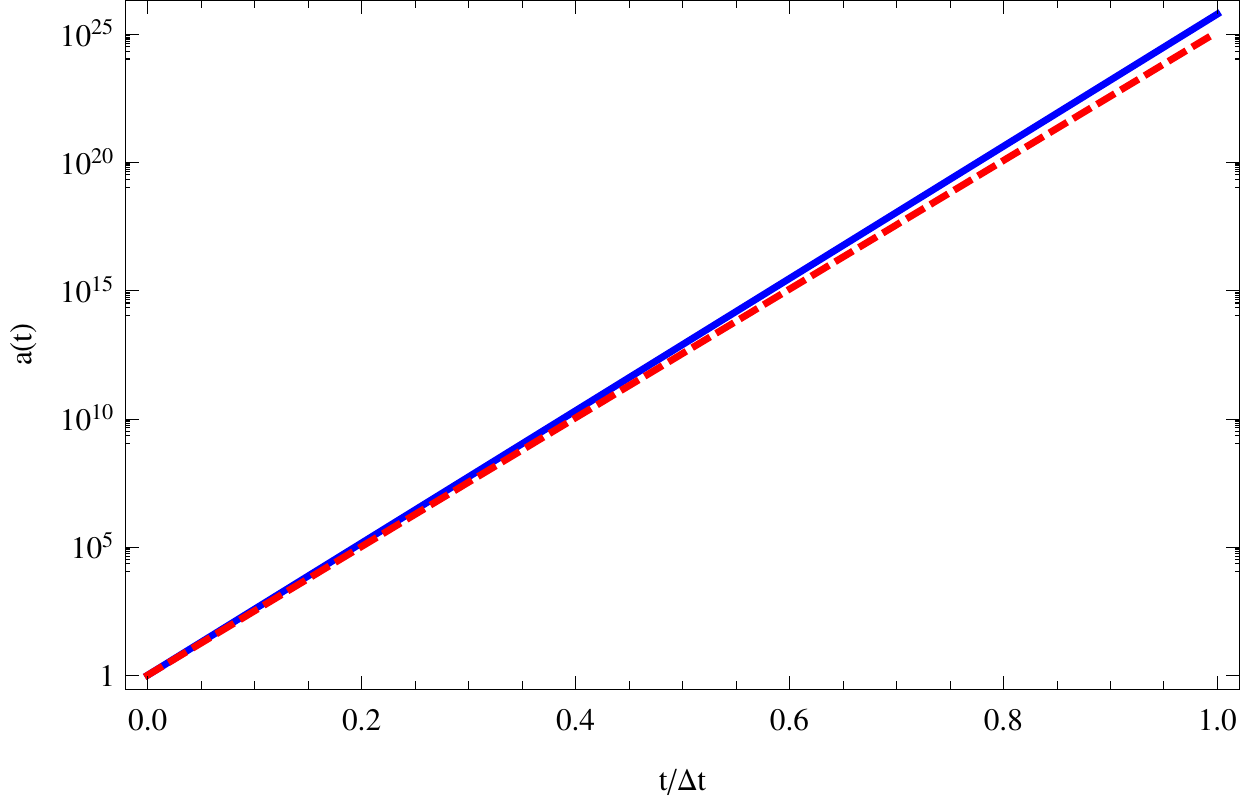}

\caption{The coupled differential equations (\ref{b12a}) and (\ref{b12b}) has no 
exact analytical solutions. However they can be evaluated numerically. In this figure we 
compare the numerical solution regarding evolution of the radion field with 
the theoretical solution as presented in Eq. (\ref{bm10}). We have scaled the time unit such 
that inflationary time scale becomes equivalent to the range $[0,1]$, which 
is accomplished by dividing time $t$ with $\Delta t$ from Eq. (\ref{TimeRad}). Also 
the radion field is normalized with respect to 
$r_{0}$ which is also in the Planck scale. Along with 
we have also presented the numerical evolution of the scale factor 
compared to the theoretical one presented through Eq. (\ref{bm4a}) with 
dotted one being the theoretical solution and solid line represents 
numerical solution. Note that both the scale factor shows an e-folding 
of order 60 as desired. The good match between these two curves in respective units 
show the validity 
of our result and the justification of using the small time approximation. \label{fig01}}

\end{center}
\end{figure}
%%%%%%%%%%%%%%%%%%%%%%%%%%%%%%%%%%%%%%%%%%%%%%%%%%%%%%%%%%%%%%%%%%%%%%%%%%%%%%%%%
%%%%%%%%%%%%%%%%%%%%%%%%%%%%%%%%%%%%%%%%%%%%%%%%%%%%%%%%%%%%%%%%%%%%%%%%%%%%%%

Thus the radion field decreasing with time trigger the inflation which takes over the 
brane is in complete agreement with recent theoretical and experimental observations 
\cite{Wise2}. We should also mention in this context that the inflation 
can be thought to be driven by the potential generated by the bulk scalar field 
which stabilizes the radion field as described in the next section. As the radion 
field runs to its stabilized value, the bulk scalar also goes toward the minima 
of the potential and thus triggers the inflation. As the radion field gets 
stabilized the bulk scalar resides at the minima of the potential halting the inflation.
Then the evolution of the radion field could be given by equation (\ref{bm10}). 

One thing has to be mentioned in this context, in inflation an important parameter 
is the slow roll parameter determining the departure from the exponential 
expansion. For an ideal situation the slow roll parameter should vanish. 
In general for all inflationary scenario (with very few exceptions) the slow roll 
parameter is negligible. In general the slow roll parameter is defined by, 
$\eta _{H}=-\ddot{H}/2H\dot{H}$, which for our case leads to, 
$\frac{\dddot{r}}{\ddot{r}\dot{r}}\frac{1-\Omega _{c}^{2}}{2k\pi \Omega _{c}^{2}}$. 
This exactly vanishes for our radion field evolution given by Eq. (\ref{bm10}). However 
from Figure \ref{fig01} it is clear that there is a very tiny difference between 
the theoretical curve and numerical curve suggesting that even if the $\dddot{r}$ 
term exists it would be extremely small and hence the slow roll parameter would 
satisfy $\eta _{H}\ll 1$.

Thus after the inflationary phase is over, the radion field would have some small value 
with which it will remain forever. Hence we have considered Einstein's equation with a 
specific form for the metric ansatz which is particularly suitable for examining the 
scale factor variation with time. We have considered the dynamics of the radion field 
showing that the radion field gets evolved by the energy density on the visible brane 
which in turn makes the scale factor to develop. However as the radion field gets 
stabilized and decreases in value at late times, the scale factor variation will be 
governed by the energy density alone leading to standard cosmology.
In order to get that particular value in terms of physical parameters of our system 
we need to find a stabilization mechanism, which we will address next.

%%%%%%%%%%%%%%%%%%%%%%%%%%%%%%%%%%%%%%%%%%%%%%
%%%%%%%%%%%%%%%%%%%%%%%%%%%%%%%%%%%%%%%%%%%%%
\section{Dynamics of Radion Stabilization}\label{branerad}

For the scenario as depicted in the work of Randall and Sundrum
(see \cite{Randall1}), the radion field is associated with the Vacuum Expectation Value(VEV) 
of a four dimensional massless scalar field that has zero potential and
it's VEV is  not determined by the dynamics of the model. Thus it
was necessary to determine the mechanism to stabilize the radion
field. This work was done by Goldberger and Wise (see
\cite{Wise2}) for time independent radion field, using a bulk
scalar field with interaction term localized on 3-branes. 
Hence the derivation was completely classical. They
have taken the bulk scalar field $\Phi$ to depend on the extra
space dimension $\phi$ and have obtained $kr_{c}\sim 12$ without
any fine tuning of the parameters. 

In the context of string theory where one encounters several moduli, 
such stabilization has been addressed by the presence of various antisymmetric tensor 
fields in the bulk spacetime. In particular, the Klebanov-Strassler 
throat geometry with a $D_{3}$-$D_{7}$ brane configuration has a 
close resemblance with warped geometry and can be stabilized by 
3-form fluxes \cite{Kodama,Kachru}. All geometric 
moduli can thus be classically stabilized by the field strengths of 
various form fields. In the Type IIA theory AdS4 vacua 
can be realized in terms of branes, which provide solutions between AdS4 
and 4-dimensional Minkowski spacetime inducing transitions between different vacua.
An interesting aspect of the type II theory is that they are 
also dynamically unstable in the moduli sector. This represents the fact that 
scale of internal space can be fixed for a quite 
long time in a region with a large warp factor \cite{Gibbons,DeWolfe2,Font}.
Here in an effort to capture the dynamics of moduli stabilization mechanism in braneworld scenario 
we present a simple time dependent generalizations of Goldberger-Wise 
mechanism with a scalar field in the bulk. Using time dependent bulk scalar 
field we have addressed the dynamics of moduli stabilization to 
determine the evolution of the modulus to its stable value which 
also resolves the gauge hierarchy problem. We further relate 
this dynamical stabilization with the inflationary model 
of the universe. 

The total action for the time dependent scalar field is given by,
%%%%%%%%%%%%%%%%%%%%%%%%%%%%%%%%%%%%%%%%5
\begin{equation}\label{b18}
S_{bulk}=\frac{1}{2}\int d^{4}x \int _{-\pi}^{\pi} d\phi \sqrt{-G} 
\left[G^{AB}\partial _{A}\Phi \partial _{B} \Phi + m^{2} \Phi ^{2}\right]
\end{equation}
%%%%%%%%%%%%%%%%%%%%%%%%%%%%%%%%%
where $G_{AB}$ with $A,B=\mu , \phi$ is given by equation ($\ref{b1}$).
We also include the interaction terms on the hidden and visible branes 
(at $\phi =0$ and $\phi =\pi$, respectively) as,
%%%%%%%%%%%%%%%%%%%%%%%%%%%%%%%%%%%%%%%
\begin{equation}\label{b19}
S_{planck}=\int d^{4}x \int _{-\pi}^{\pi} d\phi \sqrt{-g_{h}}\lambda _{h}
\left[\Phi ^{2}-V_{h}^{2} \right]^{2}\delta (\phi - 0),
\end{equation}
%%%%%%%%%%%%%%%%%%%%%%%%%%%%%%%%%%
and,
%%%%%%%%%%%%%%%%%%%%%%%%%%%%%%%%%%%
\begin{equation}\label{b20}
S_{visible}=\int d^{4}x \int _{-\pi}^{\pi} d\phi \sqrt{-g_{v}}\lambda _{v}
\left[\Phi ^{2}-V_{v}^{2} \right]^{2}\delta (\phi - \pi),
\end{equation}
%%%%%%%%%%%%%%%%%%%%%%%%%%%%%%%%%%%%%%%%
where $g_{h}$ and $g_{v}$ are the determinants of the induced metric on the hidden and visible branes, 
respectively. $\Phi (\phi ,t)$ can be determined by solving the following differential equation,
%%%%%%%%%%%%%%%%%%%%%%%%%%%%%%%%%%%%%%%%%%
\begin{eqnarray}\label{b21}
0&=&-\frac{\partial}{\partial \phi} \left[ \frac{e^{-4A}a^{3}}{r} \partial _{\phi} \Phi \right] +
\frac{\partial}{\partial t} \left[ e^{-2A}a^{3}r \partial _{t}\Phi \right]
\nonumber
\\
&+&m^{2}e^{-4A}a^{3}r\Phi +4e^{-4A}a^{3}\lambda _{h}\Phi \left( \Phi ^{2}-V_{h}^{2} \right)\delta (\phi)
+4e^{-4A}a^{3}\lambda _{v}\Phi \left( \Phi ^{2}-V_{v}^{2} \right)\delta (\phi - \pi)
\end{eqnarray}
%%%%%%%%%%%%%%%%%%%%%%%%%%%%%%%%%%%%
Choosing the solution as,
%%%%%%%%%%%%%%%%%%%%%%%%%%%%%%%%%%%%%%5
\begin{equation}\label{b22}
\Phi (\phi ,t)=T(t)e^{2A\left(\phi ,t\right)}\left[X(t)e^{\nu A\left(\phi ,t\right)}+Y(t)e^{-\nu A\left( \phi ,t\right)} \right]
\end{equation}
%%%%%%%%%%%%%%%%%%%%%%%%%%%%
where $A\left(\phi ,t\right)=k\mid \phi \mid r(t)$. 
Then the time dependent part of the differential equation reduces to,
%%%%%%%%%%%%%%%%%%%%%%%%%%%%%%%%%%%%55
\begin{equation}\label{b23}
a^{3}r\left[e^{\nu A}\left\lbrace TX_{t}+X\left(T_{t}+(2+\nu)TA_{t}\right)\right\rbrace 
+e^{-\nu A}\left\lbrace TY_{t}+
Y\left(T_{t}+(2-\nu)TA_{t}\right)\right\rbrace \right]=C(\phi)
\end{equation}
%%%%%%%%%%%%%%%%%%%%%%%%%%%%%%%%%%%%
where $X_{t}$, $Y_{t}$, $A_{t}$ and $T_{t}$ denote time derivatives of the respective functions 
$x$, $Y$, $A$ and $T$. $C(\phi)$ is a $\phi$ dependent integration constant. 
The ansatz for the variables $X(t)$ and $Y(t)$ is given by,
%%%%%%%%%%%%%%%%%%%%%%%%%%%%%%%%%%
\begin{eqnarray}\label{b23a}
X(t)=v_{v}e^{-(2+\nu)k\pi r(t)}-v_{h}e^{-2\nu k\pi r(t)}
\nonumber
\\
Y(t)=v_{h}\left(1+e^{-2\nu k\pi r(t)}\right)-v_{v}e^{-(2+\nu)k\pi r(t)}
\end{eqnarray}
%%%%%%%%%%%%%%%%%%%%%%%%%%%%%%%%%%%%%
This choice is motivated from the Goldberger-Wise stabilization mechanism, where these two functions 
$X$ and $Y$ were time independent. We have taken this choice such that our solution can be mapped 
to GW solution for time independent radion field scenario quiet easily. 
With this choice we can determine the two quantities $v_{v}$ and $v_{h}$ giving 
the respective boundary conditions such that,
%%%%%%%%%%%%%%%%%%%%%%%%%%%%%%%%%%%
\begin{eqnarray}\label{b23b}
\Phi(0,t)=T(t)v_{h}
\nonumber
\\
\Phi(\pi ,t)=T(t)v_{v}
\end{eqnarray}
%%%%%%%%%%%%%%%%%%%%%%%%%%%%%%%%%%%%%
The above expression now brings out the physical meaning of the function $T(t)$. The 
function represents the departure of the GW solution due to inclusion of 
time dependence to the radion field. It also determines the scalar field values at 
the boundary points.
Using the forms of $X(t)$ and $Y(t)$ from equation (\ref{b23a}) and the form of 
$a(t)$ and $r(t)$ from equations (\ref{bm4a}) and (\ref{bm4}), the function $T(t)$ 
is determined as, $T(t) \propto \left(1-e^{-2\kappa t}\right)$.
Thus we see that at large time our solution reduces to that of Goldberger and Wise 
solution. Finally solving for the potential and then minimizing it we readily obtain,
%%%%%%%%%%%%%%%%%%%%%%%%%%%%%%%%%
\begin{equation}\label{b24}
r(t)=\left(\frac{4}{\pi}\right)\frac{k^{2}}{m^{2}}\ln\left(\frac{v_{h}}{v_{v}}\right) 
+ \mathcal{O} \left(e^{-2\kappa t}\right)
\end{equation}
%%%%%%%%%%%%%%%%%%%%%%%%%%%%%%%%
Hence the radion field has the same stabilized value as predicted by Goldberger and Wise \cite{Wise2}, 
with an extra correction factor which decay exponentially 
with time and decreases to such a small value that the first term only contributes 
to the stabilized value for the radion field. Hence the time dependent radion field can 
be stabilized by introducing time dependent scalar field in the bulk. By minimizing the 
potential due to the scalar field we readily obtain the stabilized value which is exactly 
the GW value at late times. Due to this stabilization, the cosmic evolution exits from 
inflation and follow standard cosmology at late times.
%%%%%%%%%%%%%%%%%%%%%%%%%%%%%%%%%%%%%%%%%%%%%%%
%%%%%%%%%%%%%%%%%%%%%%%%%%%%%%%%%%%%%%%%%%%%%%

\section{Discussion}

In this work we have generalized the RS model for time dependent radion field and have studied the 
dynamics of the moduli. The constraint between matter energy density in visible and hidden 
brane appears as a consequence of requiring static solution without stabilizing it. This constraint 
never appear in a dynamical theory as considered in this work. We have further considered the 
evolution of scale factor as determined from the evolution of the radion field, which in turn 
is determined by the energy density on the visible brane. The equivalence of this approach with covariant 
curvature formalism has also been addressed. We have then shown that the inflationary 
epoch can be connected with the radion field getting compactified and stabilized to a small value. 
Thus according to our model this decrease of the moduli can trigger the inflation in visible brane. 
This seems a very natural and interesting scenario for inflation on the visible brane in the RS 
brane world. Also note that the evolution of the scale factor or radion field is never affected 
by the bulk cosmological constant which is due to the reason that in RS model there is a fine 
tuning between brane tension and bulk cosmological constant. This also comes out quiet naturally 
from our calculations.
In order to explain the exit from inflation  we need to 
know the stabilization of the dynamical radion field. 
For this purpose we have discussed how this radion field gets stabilized to the value obtained 
by Goldberger and Wise by introducing a time dependent bulk scalar field. The dynamics of the moduli 
can be determined by working out the potential for this time dependent bulk scalar field and 
then through the potential we could determine the stabilized value of bulk scalar field which 
coincides with the GW value. This shows that the method of Goldberger and Wise works for time 
dependent case as well which we have generalized from their original time independent scenario. 
The time dependent part appears to decrease exponentially with time and thus have no relevance in the 
present epoch. This in turn explains how the issue of gauge hierarchy problem in connection with the mass 
of the Higgs boson in standard model can also be resolved 
in such an warped geometry model where the radion 
field needs to be  stabilized to a small value $\sim$ inverse Planck length, as required by the RS model.

%%%%%%%%%%%%%%%%%%%%%%%%%%%%%%%%%%%%%%%%%%%%%%%%%
%%%%%%%%%%%%%%%%%%%%%%%%%%%%%%%%%%%%%%%%%%%%%%%%%

\section*{Acknowledgements}

S.C. is funded by a SPM Fellowship from CSIR, Government of India. The authors would 
also like to thank the reviewer for helping to improve the manuscript.  

%%%%%%%%%%%%%%%%%%%%%%%%%%%%%%%%%%%%%%%%%%%%%%%%%%%%%%%%%%%%
%%%%%%%%%%%%%%%%%%%%%%%%%%%%%%%%%%%%%%%%

\end{document}